%% file: icassp2025-asr_llm_contextualization.tex
\documentclass[conference]{IEEEtran}
\IEEEoverridecommandlockouts

\usepackage{cite}
\usepackage{amsmath,amssymb,amsfonts}
\usepackage{algorithmic}
\usepackage{graphicx}
\usepackage{textcomp}
\usepackage{xcolor}
\usepackage{multirow}
\usepackage{booktabs}
\usepackage[inline]{enumitem}

\def\BibTeX{{\rm B\kern-.05em{\sc i\kern-.025em b}\kern-.08em
    T\kern-.1667em\lower.7ex\hbox{E}\kern-.125emX}}
\begin{document}

\title{Contextualization of ASR with LLM using phonetic retrieval-based augmentation}

\author{\IEEEauthorblockN{Zhihong Lei, Xingyu Na, Mingbin Xu, Ernest Pusateri, Christophe Van Gysel, \\
Yuanyuan Zhang, Shiyi Han\textsuperscript{*}\thanks{\textsuperscript{*}work performed while at Apple} and Zhen Huang}
\IEEEauthorblockA{Apple\\
\{zlei, na\_xingyu, mingbinxu, epusateri, cvangysel\}@apple.com}}


\maketitle

\newcommand{\TODOcvangysel}[1]{\textcolor{red}{TODO(cvangysel): #1}}

\begin{abstract}
Large language models (LLMs) have shown superb capability of modeling multimodal signals including audio and text, allowing the model to generate spoken or textual response given a speech input. However, it remains a challenge for the model to recognize personal named entities, such as contacts in a phone book, when the input modality is speech. In this work, we start with a speech recognition task and propose a retrieval-based solution to contextualize the LLM: we first let the LLM detect named entities in speech without any context, then use this named entity as a query to retrieve phonetically similar named entities from a personal database and feed them to the LLM, and finally run context-aware LLM decoding. In a voice assistant task, our solution achieved up to 30.2\% relative word error rate reduction and 73.6\% relative named entity error rate reduction compared to a baseline system without contextualization. Notably, our solution by design avoids prompting the LLM with the full named entity database, making it highly efficient and applicable to large named entity databases.
\end{abstract}

\input{01-introduction}
\input{02-methodology}
\input{03-experiments}
\input{04-conclusions}

\bibliographystyle{IEEEtran}
\bibliography{icassp2025-asr_llm_contextualization}

\end{document}

%% file: 01-introduction.tex
\section{Introduction}
\label{sec:intro}

Large Language Models (LLMs) are usually trained on text input and output, but they can also be effectively adapted to accept and produce other modalities, i.e. to be multimodal.  In this work we are interested in tasks with an audio input modality like speech-to-text translation, automatic speech recognition (ASR), natural language understanding (NLU) and conversational AI  \cite{audiopalm,prompt-llm-with-asr-task, integrating-pt-speech-lm, yu2024connecting,ChatGPT4o}.  While multimodal LLMs excel at these tasks in general, they can falter for applications that require incorporating knowledge of personal context.  For instance, performing NLU for a voice assistant application benefits greatly from an awareness of a user’s contact book, but multimodal LLMs lack an efficient mechanism for incorporating such context.

While there are many different tasks that use audio input, a large number of them either explicitly or implicitly involve performing ASR, and so in this work we focus on an ASR task \cite{chen2023hyporadise} in the belief that the results should generalize to other tasks and domains with audio input. More specifically, we focus on ASR in the voice assistant domain, where recognizing personal named entities correctly is especially important.

In conventional ASR systems, various solutions based on contextual biasing have been studied to improve recognition of personal named entities \cite{lei2024personalization, deep-context, bleeker23_interspeech}. %
While achieving remarkable improvement, these solutions are not directly applicable to LLMs in a decoder-only setting due to their use of cross-attention \cite{deep-context, bleeker23_interspeech} or finite state transducers \cite{lei2024personalization, deep-context}. %
Recently, multiple works have investigated contextualization of LLM-ASR systems via prompting \cite{lakomkin2024contextualization, chen2024contextualization, li2023prompt}. Specifically, personal named entities serve as a prompt to contextualize the LLM. In practical applications such as voice assistant, however, users can have hundreds or thousands of named entities of different categories. %
%
%
Using a prompt with all personal named entities can lead to a significant increase in computational complexity and memory footprint, especially for on-device applications.

To address this challenge, we propose a phonetic retrieval-based solution to contextualize multimodal LLMs. We first let the LLM detect personal named entities in the speech, then retrieve phonetically similar named entities from the database, and, lastly, feed the retrieved named entities to the LLM for further context-aware decoding. %
This solution proves to be highly effective: up to 30.2\% word error rate reduction, and more importantly, up to 73.6\% named entity error rate reduction are achieved as compared to a baseline without contextualization. Our solution by design avoids prompting the LLM with the full named entity database, making it highly efficient and applicable to large named entity databases.

%% file: 02-methodology.tex
\newcommand{\MethodFullFull}{\textsc{full-full}}
\newcommand{\MethodNeFull}{\textsc{ne-full}}
\newcommand{\MethodFullNe}{\textsc{full-ne}}

\section{Methodology}
\label{sec:method}

\subsection{LLM-based ASR}
\label{subsec:baseline}

\begin{figure}[t!]
    \centerline{\includegraphics[width=0.7\columnwidth]{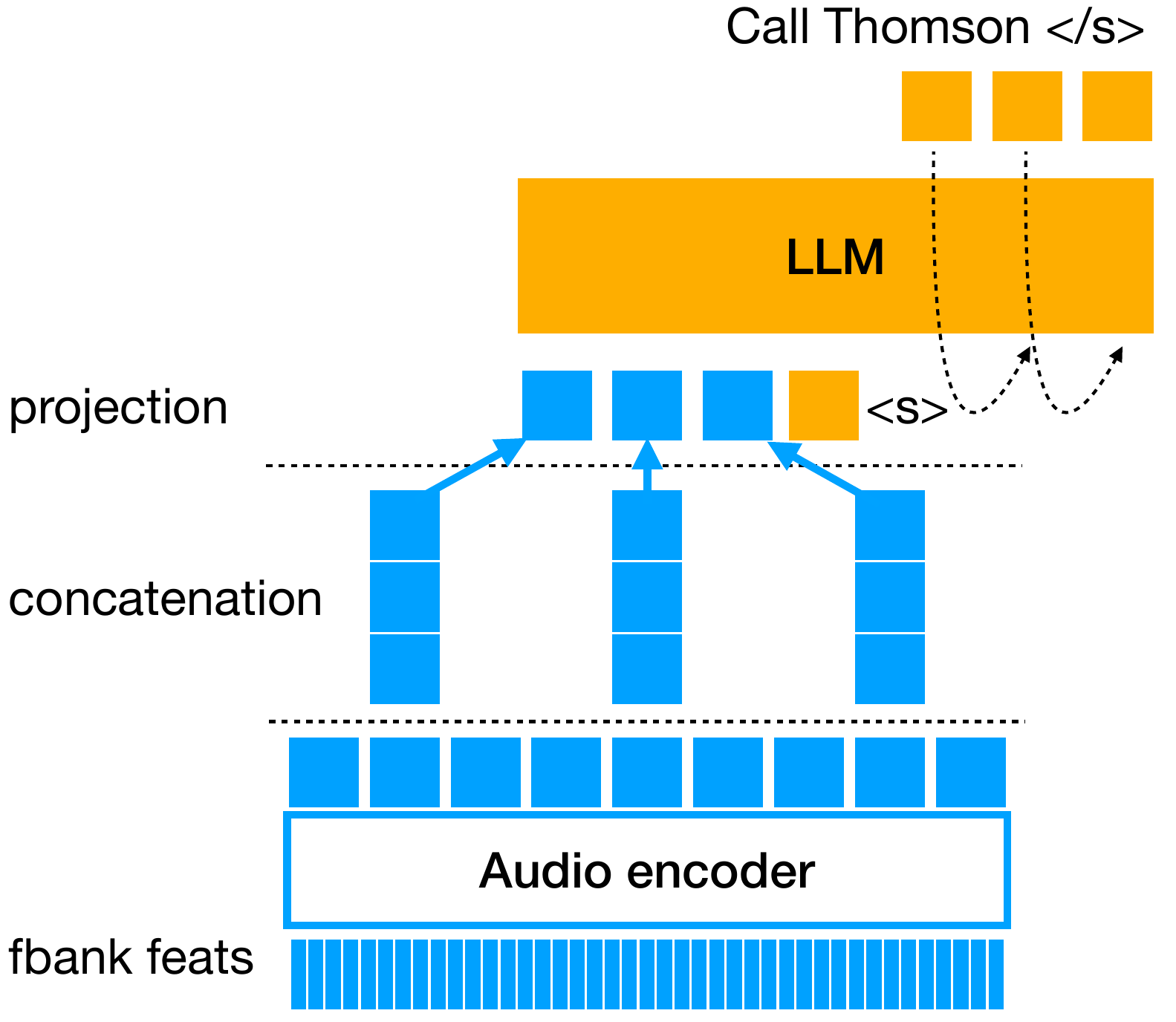}}
    \caption{Baseline LLM-ASR system. Audio features from a pretrained audio encoder are subsampled, projected and fed to the LLM for decoding.}
    \label{fig:baseline}
\end{figure}

Following \cite{yu2024connecting}, we integrate a pretrained audio encoder with a pretrained LLM (Fig.~\ref{fig:baseline}). The audio embeddings generated from the audio encoder are subsampled by concatenating every N consecutive embeddings. This subsampling has two effects: to reduce the input sequence length for faster training/inference and lower memory footprint, and to better match the frame rate of the audio embeddings to the rate of the text tokens. The subsampled audio embeddings are then projected to the embedding space of the LLM before being provided to the LLM. The LLM is finetuned on the transcripts. The sentence begin and sentence end tokens are introduced to indicate start and end of the transcript. At decoding time, when the sentence end token is generated, the decoding stops. 

\subsection{Our method}
\label{subsec:our_method}

Our contextualization method is inspired by retrieval augmented generation (RAG) \cite{lewis2020retrieval}: we retrieve and provide only relevant named entities in personal databases to the LLM. However, as compared to vanilla RAG used in text-only LLMs, we face the following challenges: %
\begin{enumerate*}[label=(\arabic*)]
    \item The input query is speech, while the named entity database is text.
    \item The named entities do not carry significant semantic differences and differ only in phonetics.
    \item The correct span in the speech input signal has to be identified as the query. %
\end{enumerate*} %
Additionally, different types of named entities, such as contact, app name and playlist, should be considered. We propose a three-step method to address these challenges: %
\begin{enumerate*}[label=(\alph*)]
    \item during \textbf{detection}, context-free ASR decoding detects named entity spans in the speech signal, followed by
    \item \textbf{retrieval} of phonetically similar named entities, and finally,
    \item we perform \textbf{generation} by running context-aware ASR decoding to generate the correct named entities.
\end{enumerate*} %

In the \textbf{detection} step, we feed the audio embeddings to the LLM and run context-free ASR decoding. To enable named entity detection, named entities are tagged in the transcripts of the training data with associated class labels, such as ``\textit{Call \textless{}contact\textgreater{} Thomson \textless{}/contact\textgreater}''. At decoding time, spans that correspond to possible named entity recognitions in the ASR hypothesis will be wrapped with tags, if detected, so they can be extracted by simply parsing the hypothesis. If no named entities are detected, decoding stops. Otherwise, we go to the \textbf{retrieval} step, where we retrieve named entities from the database that are phonetically similar to the detected named entities. The database contains different classes of named entities, each class corresponding to a flat list of named entities and their phonetic representations. In this work, we use pronunciations as the phonetic representations obtained either from a pronunciation lexicon or a grapheme-to-phoneme (G2P) model.

\begin{figure}[t!]
    \centerline{ \includegraphics[width=0.9\columnwidth]{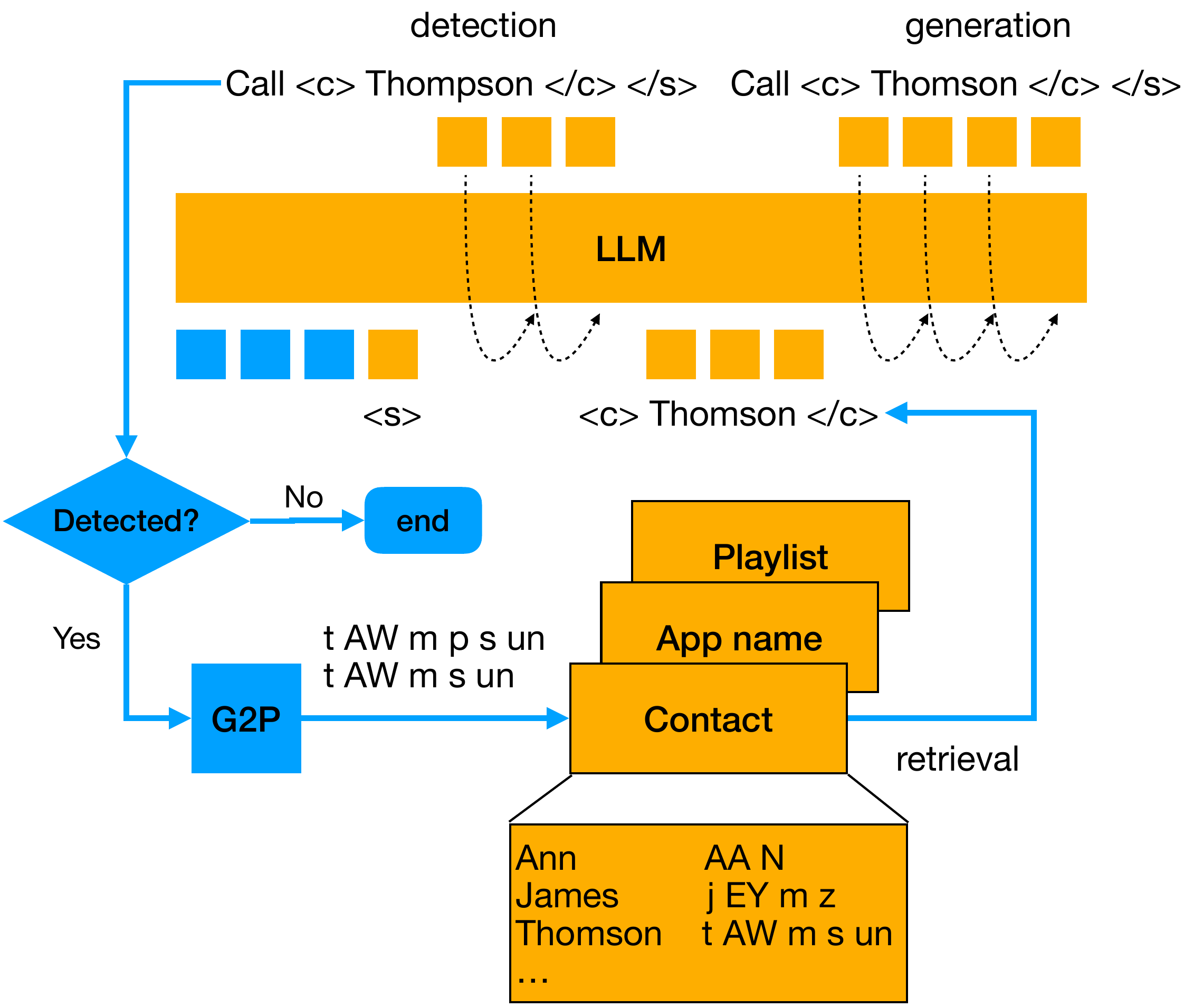}}
    \caption{Our three-step method: named entity \textbf{detection}, phonetic-based \textbf{retrieval} and context-aware \textbf{generation}. In the example above, \textless c\textgreater\ and \textless /c\textgreater\ indicate the start and the end of a \textbf{\textit{c}}ontact named entity.}
    \label{fig:method}
\end{figure}

To measure the similarity between two pronunciations, we use normalized phonetic distance (NPD). To compute NPD, we first compute the edit distance between phonetic representations of the query and database entry, then divide by the number of phonemes in the query pronunciation. Empirically, we use a simple rule to determine which named entities to be included in the prompt: the NPD is no greater than 1.2 times the best NDP or it is lower than 0.2. %
We keep a maximum of 10 candidate names in the prompt, sorted by NPD. Finally, in the \textbf{generation} step, we prompt the LLM with the retrieved named entities and run context-aware ASR decoding. This method is illusrated in Figure \ref{fig:method}.

In the proposed method, which we refer to as \MethodFullFull{} for the remainder of this paper, full ASR decoding is run twice.  We derive more efficient variations of the method by changing either the \textbf{detection} or the \textbf{generation} step as follows: %
\begin{enumerate*}[label=(\arabic*)]
    \item Within the \MethodNeFull{} method, during the \textbf{detection} step, the LLM directly identifies named entities without having to generate the full ASR hypothesis. For example, given an audio with ground-truth transcript ``\textit{Call Thomson}'', we expect the LLM to generate ``\textit{\textless contact\textgreater\ Thomson \textless /contact\textgreater}'' or another similar sounding name surrounded by tags, as the recognition hypothesis. %
    In the \textbf{generation} step, we run full ASR decoding with the context. If no named entities are detected in the \textbf{detection} step, we skip the \textbf{retrieval} step, and feed an empty context for \textbf{generation}.
    \item For the \MethodFullNe{} method, the \textbf{detection} step consists of the LLM generating the full ASR hypothesis with tagged named entities, if detected. In the \textbf{generation} step, the LLM generates only the correct named entities. We then replace named entities in the ASR hypothesis with corresponding named entities generated with context. 
\end{enumerate*} %

The three models we present are shown in Table \ref{tab:method}, producing either full ASR hypothesis or only named entities in the \textbf{detection} or the \textbf{generation} step. All three models are highly efficient at inference time. When no named entities are detected, all models reduce to a vanilla ASR model without the need of a contextual prompt. When a named entity is detected, only retrieved named entities instead of the full database are fed to the LLM as the contextual prompt, requiring a much shorter KV cache. Additionally, as compared to \textsc{full-full}, the \textsc{full-ne} and \textsc{ne-full} models further improve efficiency by reducing the number of tokens to generate.

\begin{table}[htbp]
\caption{Models}
\begin{center}
\begin{tabular}{l|cc}
    \toprule
    \multirow{2}{*}{Model} & \textbf{detection} & \textbf{generation} \\
    & (context-free) & (context-aware) \\
    \hline
    \MethodFullFull{} & ASR decoding & ASR re-decoding \\
    \MethodNeFull{} & NE detection & ASR decoding \\
    \MethodFullNe{} & ASR decoding & NE rewriting \\
    \bottomrule
    \end{tabular}
\label{tab:method}
\end{center}
\end{table}


\subsection{Training process}
\label{subsec:training_process}
We aim to train a single model for both the \textbf{detection} and \textbf{generation} steps. This is particularly challenging because the \textbf{generation} input depends on the \textbf{detection} output.  Here we will describe how we constructed training data to overcome that challenge. We will focus on constructing source sequences, as we can easily construct a corresponding target sequence from a source sequence.

We first consider \MethodFullFull{} training for audios with detected entities. In this case, to build a source sequence, we need audio features, the desired \textbf{detection} output, the expected \textbf{retrieval} output and the desired \textbf{generation} output. Audio features can be generated from audio, and it is sensible to use the tagged reference as the \textbf{generation} output.

However, it is less clear what should be used as the \textbf{detection} and \textbf{retrieval} outputs. If we used the tagged reference as the \textbf{detection} output and used tagged reference entities to generate the \textbf{retrieval} output, this would create a severe training/inference mismatch.  In fact, it is exactly the cases where the \textbf{detection} output does not match the reference where we expect to benefit from \textbf{retrieval} and \textbf{generation}. So, instead, we train a standalone detection model whose output is used as the \textbf{detection} output for training.  The standalone model output is also used to obtain the query to obtain the \textbf{retrieval} output. This gives us all the components necessary to generate source sequences for \MethodFullFull{} for audios with detected entities. For audios without detected entities, the source sequence is constructed only from the audio features and reference.  This reduces training and inference computation for audios without entities. We use ``\textless s\textgreater{}'' and ``\textless /s\textgreater{}'' to delineate different
regions of the source sequence, as shown in Figure 3.

The \MethodFullNe{} process differs from the \MethodFullFull{} process in that only the tagged entities are used as the \textbf{generation} output. The \MethodNeFull{} process differs from the \MethodFullFull{} process in two ways. First, only the tagged entity from the standalone detection model is used as the \textbf{detection} output for audios where entities are detected. Second, \textbf{generation} output is included whether or not entities have been detected.

\begin{figure}[t!]
    \centerline{\includegraphics[width=\columnwidth]{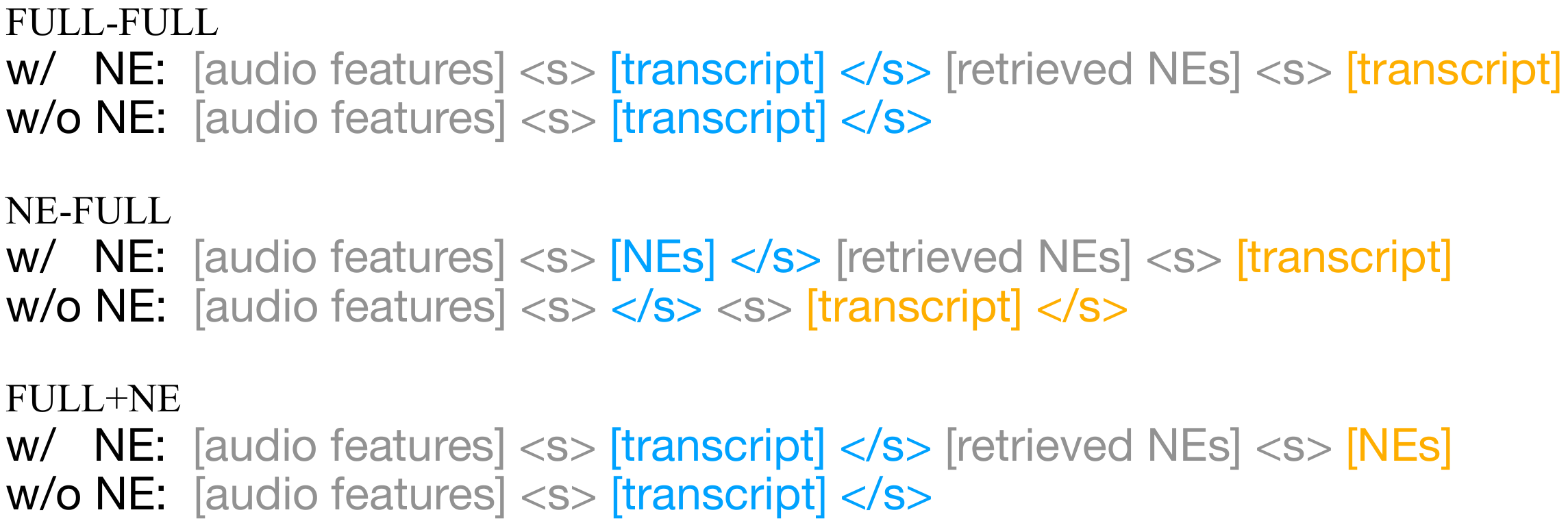}}
    \caption{Example training source sequences for audios with and without detected named entities. Regions in grey serve as prompts to the LLM. Regions in blue and yellow serve as training targets for \textbf{detection} and \textbf{generation}.}
    \label{fig:training_process}
\end{figure}

%% file: 03-experiments.tex
\section{Experiments}
\label{sec:experiment}
\subsection{Data and evaluation metrics}
\label{subsec:data}
Experiments are conducted on a voice assistant task. The training data consists of 31k hours of English data. While the ground-truth transcriptions contain tagged personal named entities, the audios are not associated with personal databases. Hence, we synthesize personal databases for our training data as follows. We extract a pool of named entities of different classes and their distributions from our training data, denoted as $NE(c)$, where $c$ indicates the class. We also estimate a distribution of the number of named entities in each class from a development set where the named entity databases are available, denoted as $N(c)$. For each audio in the training data, we synthesize named entity databases in the following way: for each class $c$, we generate a number $n(c)$ from the distribution $N(c)$, with $n(c)$ denoting the number of named entities to synthesize for the class $c$. Then, we randomly pick $n(c)$ unique named entities from the distribution $NE(c)$. Finally, if the transcript of the audio contains a named entity, we add it to the synthesized database.

Our models are evaluated on a test set containing 37k queries, each with a personal named entity database of various classes including contact, music playlist and app name. Contact stands out as the most prominent class, with 12k contacts found in 11.2k queries. On average, each contact list has hundreds of contacts. We only consider contact entities during our experiments. The models are evaluated with two metrics: word error rate (WER) and entity error rate (NER). WER provides an overall accuracy metric, while NER measures the contextualization method with higher resolution. Named entity tags are ignored when computing WER.

\subsection{Models}
\label{subsec:models}
We start with a Conformer \cite{DBLP:conf/interspeech/GulatiQCPZYHWZW20} audio encoder trained with CTC criteria \cite{DBLP:conf/icml/GravesFGS06} following \cite{lei2024personalization, lei2023acoustic, xu2023conformer}. This encoder has a 4x subsampling convolution block, 12 Conformer layers of 512 output units each, amounting to 82M parameters. The audio encoder consumes filter bank features and generates audio embeddings at a frame rate of 25 per second. The encoder is first pretrained on 300k hours of data with machine-generated transcripts, then finetuned on 45k hours supervised data. Given that the encoder has seen much more data, we consider it a reasonable baseline to compare with the LLM models.

In this work, we use the Mistral 7B pretrained LLM \cite{jiang2023mistral} for experiments. We freeze the audio encoder in LLM finetuning, use a subsampling of 12x for the audio embeddings, and a LoRA adapter \cite{hulora} with rank=8 and alpha=16. As a result, all LLM models we trained have 46M trainable parameters, with 21M parameters from the LoRA adapter and 25M parameters from the projection layer. 

We train a baseline LLM with this audio encoder following Fig.~\ref{fig:baseline} on the tagged transcripts. This model also serves as the standalone \textbf{detection} model for \MethodFullFull{} and \MethodFullNe{}. We train another model on tagged named entities to prepare training data for \MethodNeFull{}. Then, we train three models following Section \ref{subsec:training_process}. All LLM models are finetuned with cross-entropy loss, Adam optimizer, warmup leraning rate scheduler with linear increase and decay, peaking at 2e-4. 

\subsection{Results}
\label{subsec:results}

\newcommand{\StartOfSentenceToken}{\textit{\textless{}s\textgreater}}
\newcommand{\EndOfSentenceToken}{\textit{\textless{}/s\textgreater}}

We first evaluate all models without context to indicate the overall quality of the LLM-ASR models. Specifically for \MethodNeFull{}, the LLM is prompted with ``\StartOfSentenceToken{} \EndOfSentenceToken{}'' to trigger context-free decoding. We use beam search decoding with a beam of 50 and 10, respectively for the baseline CTC model and the LLM models. From the results in Table~\ref{tab:results} we can find that all LLM-ASR models significantly outperform the baseline Conformer-CTC model without context. This comparison confirms the effectiveness of using LLMs for ASR tasks, particularly given the aggressive 12x subsampling on audio embeddings. We find that \MethodFullFull{} and \MethodFullNe{} yield similar WER and NER without context, which are comparable to the baseline LLM system. Meanwhile, both WER and NER of \MethodNeFull{} are higher than the baseline. 

Next, we evaluate the models with contextualization. Following the practice in training, we include candidate named entities whose normalized phonetic distance (NPD) to queries is lower than 0.2, or it is no greater than 1.2 times the best candidate's NDP, and we keep a maximum of 10 candidate names in the prompt. With contextualization, all models achieved a significant WER and NER reduction as compared to the baseline LLM, with the largest reduction from \MethodFullFull{}: 30.2\% relative WER reduction, and 73.6\% relative NER reduction. For \MethodFullNe{}, we try to simply replace named entities in decoding hypothesis with the top-one retrieved named entities, without further prompting and context-aware generation, denoted as \textbf{Simple replacement}. Such simple replacement yields much worse WER and NER. This comparison verifies the effectiveness of the LLM in generating the correct name entities given the context-free hypothesis and the retrieved context.
\begin{table}[htbp]
    \caption{Main results}
    \begin{center}
    \begin{tabular}{l|cccc}
        \hline
    \multirow{2}{*}{Model} & \multicolumn{2}{c}{ w/o context } & \multicolumn{2}{c}{ w/ context } \\
     & WER & NER & WER & NER \\
    \hline
    \textsc{Conformer-CTC} & 8.30 & 44.86 & - & - \\
    \textsc{Baseline LLM} & 6.36 & 37.52 & - & - \\
    \hline
    \MethodFullFull{} & 6.34 & 37.58 & \textbf{4.44} & \textbf{9.90} \\
    \MethodNeFull{} & 6.53 & 38.46 & 4.99 & 15.20 \\
    \MethodFullNe{} & 6.35 & 37.61 & 4.46 & 11.09 \\
      - Simple replacement & - & - & 4.84 & 12.60 \\
    \hline
    \end{tabular}
    \label{tab:results}
    \end{center}
\end{table}

\begin{table}[htbp]
    \caption{Number of detected named entities}
    \begin{center}
    \begin{tabular}{l|cccc}
    \hline
      & Reference & \MethodFullFull{} & \MethodFullNe{} & 
    \MethodNeFull{} \\
    \cline{2-5}
      \#Detected NEs & 12.7k & 12.0k & 12.0k & 10.7k \\
    \hline
    \end{tabular}
    \label{tab:number_of_nes}
    \end{center}
\end{table}

\begin{table}[htbp]
    \caption{Impact of the number of retrieved named entities}
    \begin{center}
    \begin{tabular}{p{39pt}|p{13pt}p{13pt}p{13pt}p{13pt}p{13pt}p{13pt}p{13pt}p{13pt}}
        \hline
    \multirow{2}{*}{Model} & \multicolumn{2}{c}{1} & \multicolumn{2}{c}{ 5 } & \multicolumn{2}{c}{ 10 } & \multicolumn{2}{c}{ 20 }\\
     & WER & NER & WER & NER & WER & NER & WER & NER \\
    \hline
    \MethodFullFull{} & 4.57 & 12.15 & 4.44 & 10.52 & \textbf{4.42} & \textbf{10.28} & 4.45 & 10.57 \\
    \MethodNeFull{} & 5.09 & 16.77 & 5.00 & 15.35 & \textbf{5.00} & \textbf{15.30} & 5.03 & 15.72 \\
    \MethodFullNe{} & 4.60 & 12.93 & 4.50 & 11.49 & \textbf{4.48} & \textbf{11.30} & 4.50 & 11.50 \\

    \hline
    \end{tabular}
    \label{tab:results_by_context_length}
    \end{center}
\end{table}

\subsection{Analysis}
\label{subsec:analysis}

\subsubsection{Less accurate named entity detection results of \MethodNeFull{}}
Comparing \MethodNeFull{} with the other two models, both WER and NER are much higher. 
Our hypothesis is that named entity detection of \MethodNeFull{} is less accurate, leading to lower quality contextual prompt. To verify this hypothesis, we first count the number of contact named entities in the reference and the detection results of the models, shown in Table \ref{tab:number_of_nes}. 
We find that \MethodFullFull{} and \MethodFullNe{} both detected marginally more contacts than the reference, while \MethodNeFull{} detected significantly fewer contacts. We further use the detected named entities of the \MethodFullNe{} model to generate contextual prompt instead, and the \MethodNeFull{} model for ASR decoding with the prompt, leading to a significant WER and NER reduction as compared to using \MethodNeFull{} for both named entity detection and ASR decoding: WER decreased from 4.99 to 4.61, and NER from 15.20 to 10.2. These analyses confirm our hypothesis that \MethodNeFull{} yields less accurate named entity detection results. In future work, it would be interesting to improve the detection quality of the \MethodNeFull{} model.

\subsubsection{Impact of the number of retrieved named entities}
We retrieve and feed a fixed number (=1, 5, 10, 20) of phonetically similar named entities to the LLM. The results are shown in Table \ref{tab:results_by_context_length}. As the number of retrieved named entities grows from 1 to 10, all models yield lower WER and NER. When even more (=20) named entities are fed to the LLM, WER and NER start to increase due to noisier contextual prompt.

%% file: 04-conclusions.tex
\section{Conclusion}
\label{sec:conclusion}

In this work, we proposed a three-step contextualization solution to ASR with LLM: named entity detection, phonetic retrieval, and context-aware generation. In a voice assistant task, our solution achieved up to 30.2\% relative word error rate reduction, and 73.6\% relative named entity error rate reduction. The high accuracy and low inference time complexity makes our method an appealing solution to voice assistants with LLMs. One main limitation of the current approach is that we use edit distance to measure the phonetic distance between two pronunciations and do not take into account similarities between two phonemes. We believe that improved phonetic representations can lead to further improvement of the method.

%% file: icassp2025-asr_llm_contextualization.bbl
\begin{thebibliography}{10}
\providecommand{\url}[1]{#1}
\csname url@samestyle\endcsname
\providecommand{\newblock}{\relax}
\providecommand{\bibinfo}[2]{#2}
\providecommand{\BIBentrySTDinterwordspacing}{\spaceskip=0pt\relax}
\providecommand{\BIBentryALTinterwordstretchfactor}{4}
\providecommand{\BIBentryALTinterwordspacing}{\spaceskip=\fontdimen2\font plus
\BIBentryALTinterwordstretchfactor\fontdimen3\font minus \fontdimen4\font\relax}
\providecommand{\BIBforeignlanguage}[2]{{%
\expandafter\ifx\csname l@#1\endcsname\relax
\typeout{** WARNING: IEEEtran.bst: No hyphenation pattern has been}%
\typeout{** loaded for the language `#1'. Using the pattern for}%
\typeout{** the default language instead.}%
\else
\language=\csname l@#1\endcsname
\fi
#2}}
\providecommand{\BIBdecl}{\relax}
\BIBdecl

\bibitem{audiopalm}
P.~K. Rubenstein, C.~Asawaroengchai, D.~D. Nguyen, A.~Bapna, Z.~Borsos, F.~d.~C. Quitry, P.~Chen, D.~E. Badawy, W.~Han, E.~Kharitonov \emph{et~al.}, ``{AudioPaLM}: A large language model that can speak and listen,'' \emph{arXiv preprint arXiv:2306.12925}, 2023.

\bibitem{prompt-llm-with-asr-task}
Y.~Fathullah, C.~Wu, E.~Lakomkin, J.~Jia, Y.~Shangguan, K.~Li, J.~Guo, W.~Xiong, J.~Mahadeokar, O.~Kalinli, C.~Fuegen, and M.~Seltzer, ``Prompting large language models with speech recognition abilities,'' in \emph{ICASSP 2024 - 2024 IEEE International Conference on Acoustics, Speech and Signal Processing (ICASSP)}, 2024, pp. 13\,351--13\,355.

\bibitem{integrating-pt-speech-lm}
Y.~Hono, K.~Mitsuda, T.~Zhao, K.~Mitsui, T.~Wakatsuki, and K.~Sawada, ``Integrating pre-trained speech and language models for end-to-end speech recognition,'' in \emph{Findings of ACL 2024}, 2024.

\bibitem{yu2024connecting}
W.~Yu, C.~Tang, G.~Sun, X.~Chen, T.~Tan, W.~Li, L.~Lu, Z.~Ma, and C.~Zhang, ``Connecting speech encoder and large language model for {ASR},'' in \emph{2024 IEEE International Conference on Acoustics, Speech and Signal Processing (ICASSP)}.\hskip 1em plus 0.5em minus 0.4em\relax IEEE, 2024, pp. 12\,637--12\,641.

\bibitem{ChatGPT4o}
{OpenAI}, ``{ChatGPT-4 Turbo with vision},'' \url{https://chat.openai.com/}, 2023, version accessed on August 27, 2024.

\bibitem{chen2023hyporadise}
C.~Chen, Y.~Hu, C.-H.~H. Yang, S.~M. Siniscalchi, P.-Y. Chen, and E.-S. Chng, ``{HyPoradise}: An open baseline for generative speech recognition with large language models,'' in \emph{Advances in Neural Information Processing Systems}, vol.~36, 2023, pp. 31\,665--31\,688.

\bibitem{lei2024personalization}
Z.~Lei, E.~Pusateri, S.~Han, L.~Liu, M.~Xu, T.~Ng, R.~Travadi, Y.~Zhang, M.~Hannemann, M.-H. Siu \emph{et~al.}, ``Personalization of {CTC}-based end-to-end speech recognition using pronunciation-driven subword tokenization,'' in \emph{ICASSP 2024-2024 IEEE International Conference on Acoustics, Speech and Signal Processing (ICASSP)}.\hskip 1em plus 0.5em minus 0.4em\relax IEEE, 2024, pp. 10\,096--10\,100.

\bibitem{deep-context}
G.~Pundak, T.~N. Sainath, R.~Prabhavalkar, A.~Kannan, and D.~Zhao, ``Deep context: End-to-end contextual speech recognition,'' in \emph{2018 IEEE Spoken Language Technology Workshop (SLT)}, 2018, pp. 418--425.

\bibitem{bleeker23_interspeech}
M.~Bleeker, P.~Swietojanski, S.~Braun, and X.~Zhuang, ``{Approximate Nearest Neighbour Phrase Mining for Contextual Speech Recognition},'' in \emph{Proc. INTERSPEECH 2023}, 2023, pp. 939--943.

\bibitem{lakomkin2024contextualization}
E.~Lakomkin, C.~Wu, Y.~Fathullah, O.~Kalinli, M.~L. Seltzer, and C.~Fuegen, ``End-to-end speech recognition contextualization with large language models,'' in \emph{IEEE International Conference on Acoustics, Speech and Signal Processing (ICASSP)}, 2024, pp. 12\,406--12\,410.

\bibitem{chen2024contextualization}
Z.~Chen, H.~Huang, A.~Andrusenko, O.~Hrinchuk, K.~C. Puvvada, J.~Li, S.~Ghosh, J.~Balam, and B.~Ginsburg, ``{SALM}: Speech-augmented language model with in-context learning for speech recognition and translation,'' in \emph{IEEE International Conference on Acoustics, Speech and Signal Processing (ICASSP)}, 2024, pp. 13\,521--13\,525.

\bibitem{li2023prompt}
Y.~Li, Y.~Wu, J.~Li, and S.~Liu, ``Prompting large language models for zero-shot domain adaptation in speech recognition,'' in \emph{2023 IEEE Automatic Speech Recognition and Understanding Workshop (ASRU)}, 2023, pp. 1--8.

\bibitem{lewis2020retrieval}
P.~Lewis, E.~Perez, A.~Piktus, F.~Petroni, V.~Karpukhin, N.~Goyal, H.~K{\"u}ttler, M.~Lewis, W.-t. Yih, T.~Rockt{\"a}schel \emph{et~al.}, ``Retrieval-augmented generation for knowledge-intensive nlp tasks,'' \emph{Advances in Neural Information Processing Systems}, vol.~33, pp. 9459--9474, 2020.

\bibitem{DBLP:conf/interspeech/GulatiQCPZYHWZW20}
\BIBentryALTinterwordspacing
A.~Gulati, J.~Qin, C.~Chiu, N.~Parmar, Y.~Zhang, J.~Yu, W.~Han, S.~Wang, Z.~Zhang, Y.~Wu, and R.~Pang, ``Conformer: Convolution-augmented transformer for speech recognition,'' in \emph{21st Annual Conference of the International Speech Communication Association}.\hskip 1em plus 0.5em minus 0.4em\relax {ISCA}, 2020, pp. 5036--5040. [Online]. Available: \url{https://doi.org/10.21437/Interspeech.2020-3015}
\BIBentrySTDinterwordspacing

\bibitem{DBLP:conf/icml/GravesFGS06}
\BIBentryALTinterwordspacing
A.~Graves, S.~Fern{\'{a}}ndez, F.~J. Gomez, and J.~Schmidhuber, ``Connectionist temporal classification: labelling unsegmented sequence data with recurrent neural networks,'' in \emph{Proceedings of the Twenty-Third International Conference on Machine Learning {(ICML} 2006)}, vol. 148.\hskip 1em plus 0.5em minus 0.4em\relax {ACM}, 2006, pp. 369--376. [Online]. Available: \url{https://doi.org/10.1145/1143844.1143891}
\BIBentrySTDinterwordspacing

\bibitem{lei2023acoustic}
Z.~Lei, M.~Xu, S.~Han, L.~Liu, Z.~Huang, T.~Ng, Y.~Zhang, E.~Pusateri, M.~Hannemann, Y.~Deng \emph{et~al.}, ``Acoustic model fusion for end-to-end speech recognition,'' in \emph{2023 IEEE Automatic Speech Recognition and Understanding Workshop (ASRU)}.\hskip 1em plus 0.5em minus 0.4em\relax IEEE, 2023, pp. 1--7.

\bibitem{xu2023conformer}
M.~Xu, A.~Jin, S.~Wang, M.~Su, T.~Ng, H.~Mason, S.~Han, Y.~Deng, Z.~Huang, and M.~Krishnamoorthy, ``Conformer-based speech recognition on extreme edge-computing devices,'' in \emph{Proceedings of NAACL 2024}.\hskip 1em plus 0.5em minus 0.4em\relax Association for Computational Linguistics, 2024, pp. 131--139.

\bibitem{jiang2023mistral}
A.~Q. Jiang, A.~Sablayrolles, A.~Mensch, C.~Bamford, D.~S. Chaplot, D.~d.~l. Casas, F.~Bressand, G.~Lengyel, G.~Lample, L.~Saulnier \emph{et~al.}, ``Mistral {7B},'' \emph{arXiv preprint arXiv:2310.06825}, 2023.

\bibitem{hulora}
E.~J. Hu, P.~Wallis, Z.~Allen-Zhu, Y.~Li, S.~Wang, L.~Wang, W.~Chen \emph{et~al.}, ``{LoRA}: Low-rank adaptation of large language models,'' in \emph{International Conference on Learning Representations}, 2022.

\end{thebibliography}
